# Violação da desigualdade probabilística de Clauser-Horne-Shimony-Holt na Mecânica Quântica

Felipe Andrade Velozo[1]
Diogo Francisco Rossoni[2]
José Alberto Casto Nogales Vera[3]
Lucas Monteiro Chaves[4]
Devanil Jaques de Souza[5]

**Resumo:** Em 1982, Alain Aspect, e colaboradores, realizaram um experimento, para observar a violação da desigualdade de Clauser-Horne-Shimony-Holt na prática. Após realizar o experimento, usaram os dados na desigualdade e concluíram que a desigualdade, obtida por meio de argumentos probabilísticos, era violada, confirmando as conclusões obtidas por John S. Bell de que não era possível uma teoria de variáveis ocultas nas condições propostas por Einstein, Podolsky e Rosen. O objetivo deste trabalho é observar a validade das fórmulas nas condições existentes da teoria probabilística, para levar a uma conclusão condizente com os axiomas da probabilidade.

**Palavras-chave:** Violação da desigualdade de CHSH; Axiomas de Kolmogorov

## Introdução

O experimento [1] consiste em uma fonte de pares de fótons com polarizações correlacionadas, que são emitidos pela fonte e se distanciam um do outro e passam cada um por um polarizador com sentido da polarização representado pelo ângulo de orientação $\theta_1$ e $\theta_2$, respectivamente. O ângulo $\theta_k$ fornece o vetor $\vec{r}_k = \cos\theta_k \cdot \vec{\imath} + \text{sen}\,\theta_k \cdot \vec{\jmath}$, que indica a orientação da polarização. Ao passar pelo polarizador, cada fóton irá atravessá-lo ($U_k(w) = +1$), ou não ($U_k(w) = -1$), com $k=1$ para o 1º fóton e $k=2$ para o 2º fóton. A probabilidade [4] de cada fóton atravessar ou não, é dado por

$$\mathcal{P}_{U_1,U_2}\big(U_1(w), U_2(w); \theta_{1,2}\big) = \begin{cases} \dfrac{1}{2} \cdot \cos^2\theta_{1,2} \Leftarrow \big(U_1(w), U_2(w)\big) \in \{(-1,+1),(+1,-1)\} \\ \dfrac{1}{2} \cdot \text{sen}^2\,\theta_{1,2} \Leftarrow \big(U_1(w), U_2(w)\big) \in \{(-1,-1),(+1,+1)\} \end{cases}$$

em que $\theta_{1,2} = \theta_2 - \theta_1$ (diferença dos ângulos das orientações dos polarizadores) e as variáveis aleatórias $U_1$ e $U_2$ são a passagem ou não por seus respectivos polarizadores.

**Material e métodos**

Em busca de determinar o $w$ associado ao experimento de medir a travessia ou não dos fótons que tenham sido emitidos com a propriedade de polarização correlacionados, admitir-se-á mais duas orientações de polarizadores (totalizando 4 orientações: $a_1$, $a_2$, $b_1$ e $b_2$) e mais duas medidas (totalizando 4 variáveis aleatórias: $\{U_1(w), U_2(w), V_1(w), V_2(w)\} \subset \{-1; +1\}$. Portanto o experimento será feito o mesmo número de vezes para os polarizadores com os seguintes sentidos: a) Sentido $a_1$ no polarizador $I$ e $b_1$ no $II$; b) Sentido $a_1$ no polarizador $I$ e $b_2$ no $II'$; c) Sentido $a_2$ no polarizador $I'$ e $b_1$ no $II$; d) Sentido $a_2$ no polarizador $I'$ e $b_2$ no $II'$.

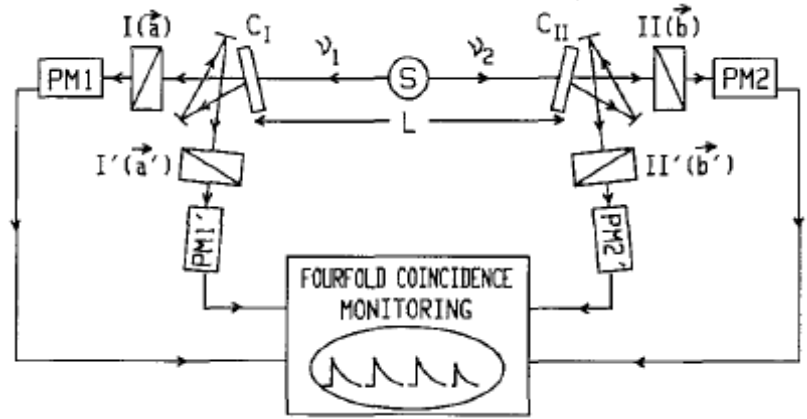

**Figura 1** Figura extraída de [1] do experimento em que: S é a fonte de fótons, $C_I$ e $C_{II}$ são instrumentos que servem para desviar os fótons ou não (tornando possível 2 percursos diferentes), em cada percurso haverá um polarizador com um determinado sentido da polarização, para o fóton 1 ($v_1$) há os polarizadores $I(\vec{a})$ e $I'(\vec{a'})$, para o fóton2 ($v_2$) há os polarizadores $II(\vec{b})$ e $II'(\vec{b'})$, PM1 e PM2 são os detectores (fotomultiplicadores)

A demonstração da desigualdade [5] se baseará apenas em argumentos presentes na Estatística, depois serão feitas comparações entre a previsão feita pela Estatística e o resultado obtido pela Mecânica Quântica. Por simplicidade, serão omitidos os parâmetros $a_j$, $b_k$ e $w$ na apresentação das fórmulas.

Calculando o valor esperado de $U_1 \cdot V_1 - U_1 \cdot V_2 + U_2 \cdot V_1 + U_2 \cdot V_2$ e observando que $-2 \leqq U_1 \cdot (V_1 - V_2) + U_2 \cdot (V_1 + V_2) \leqq 2$ (basta substituir os valores para se convencer da validade dessa desigualdade), tem-se

$$-2 \leqq \mathcal{E}_{U_1,U_2,V_1,V_2}(U_1 \cdot V_1) - \mathcal{E}_{U_1,U_2,V_1,V_2}(U_1 \cdot V_2) + \mathcal{E}_{U_1,U_2,V_1,V_2}(U_2 \cdot V_1) + \mathcal{E}_{U_1,U_2,V_1,V_2}(U_2 \cdot V_2) \leqq 2$$

Portanto

$$\left|\mathcal{E}_{U_1,V_1}(U_1 \cdot V_1) - \mathcal{E}_{U_1,V_2}(U_1 \cdot V_2) + \mathcal{E}_{U_2,V_1}(U_2 \cdot V_1) + \mathcal{E}_{U_2,V_2}(U_2 \cdot V_2)\right| \leqq 2$$

A covariância é dada por $\text{Cov}(U_j, V_k) \equiv \mathcal{E}_{U_j,V_k}(U_j \cdot V_k) - \mathcal{E}_{U_j}(U_j) \cdot \mathcal{E}_{V_k}(V_k)$, porém, observa-se que $\mathcal{E}_{U_j}(U_j) \equiv 0$, para qualquer $j$. Calculando a partir das funções de probabilidade fornecidas, tem-se

$$\mathcal{E}_{U_j,V_k}(U_j \cdot V_k) \equiv \cos\left(2 \cdot (a_j - b_k)\right) \equiv \cos(2 \cdot \phi_{j,k}), \qquad \phi_{j,k} := a_j - b_k$$

Substituindo na desigualdade, tem-se

$$|\cos(2 \cdot \phi_{1,1}) - \cos(2 \cdot \phi_{1,2}) + \cos(2 \cdot \phi_{2,1}) + \cos(2 \cdot \phi_{2,2})| \leqq 2$$

Ao escolher $a_1 = -\frac{1}{3} \cdot \pi$, $a_2 = 0$, $b_1 = \frac{1}{3} \cdot \pi$ e $b_2 = \frac{2}{3} \cdot \pi$, tem-se $\phi_{1,1} = \frac{2}{3} \cdot \pi$, $\phi_{1,2} = \pi$, $\phi_{2,1} = \frac{1}{3} \cdot \pi$, $\phi_{2,2} = \frac{2}{3} \cdot \pi$, portanto

$$\left| \underbrace{\cos(2 \cdot \phi_{1,1})}_{=-1/2} - \underbrace{\cos(2 \cdot \phi_{1,2})}_{=1} + \underbrace{\cos(2 \cdot \phi_{2,1})}_{=-1/2} + \underbrace{\cos(2 \cdot \phi_{2,2})}_{=-1/2} \right| \leqq 2$$
$$\underbrace{\phantom{x}}_{=5/2}$$

Mas $\frac{5}{2} = 2{,}5$ é maior que 2, portanto a desigualdade não é obedecida na Mecânica Quântica.

**Resultados**

Observando o experimento, observa-se que há uma probabilidade associada para cada trajeto possível, supondo que sejam independentes as trajetórias entre si, tem-se $p_1$ para o fóton 1 se desviar e $(1 - p_1)$ para não desviar, da mesma forma, para o fóton 2 tem-se $p_2$ de probabilidade para se desviar e $(1 - p_2)$ caso contrário. Portanto, pode-se associar a cada trajeto uma variável aleatória $W_k$, com $k \in \{1,2\}$ (para cada fóton). O fóton $k$ se desviar corresponde a $W_k = 2$, e o caso contrário $W_k = 1$. Dessa forma, tem-se na função de probabilidade as variáveis aleatórias referentes a passagem dos fótons pelos polarizadores $(U, V)$ e as referentes ao trajeto percorrido $(W_1, W_2)$ até chegarem aos polarizadores

$$\mathcal{P}_{U_1,U_2,V_1,V_2,W_1,W_2}(u_1, u_2, v_1, v_2, w_1, w_2) \equiv$$

$$\equiv \begin{cases} \mathcal{P}_{W_1}(w_1) \cdot \mathcal{P}_{W_2}(w_2) \cdot \mathcal{P}_{U_1,V_1}(u_1, v_1, \phi_{1,1}) \Leftarrow (w_1 = w_2 = 1) \wedge (u_2 = v_2 = 0) \\ \mathcal{P}_{W_1}(w_1) \cdot \mathcal{P}_{W_2}(w_2) \cdot \mathcal{P}_{U_1,V_2}(u_1, v_2, \phi_{1,2}) \Leftarrow (w_1 = w_2 - 1 = 1) \wedge (u_2 = v_1 = 0) \\ \mathcal{P}_{W_1}(w_1) \cdot \mathcal{P}_{W_2}(w_2) \cdot \mathcal{P}_{U_2,V_1}(u_2, v_1, \phi_{2,1}) \Leftarrow (w_1 - 1 = w_2 = 1) \wedge (u_1 = v_2 = 0) \\ \mathcal{P}_{W_1}(w_1) \cdot \mathcal{P}_{W_2}(w_2) \cdot \mathcal{P}_{U_2,V_2}(u_2, v_2, \phi_{2,2}) \Leftarrow (w_1 = w_2 = 2) \wedge (u_1 = v_1 = 0) \end{cases}$$

$$\mathcal{P}_{W_k}(w_k) \equiv \begin{cases} p_k \Leftarrow w_k = 2 \\ 1 - p_k \Leftarrow w_k = 1 \end{cases}, \qquad 0 \leq p_k \leq 1, \qquad k \in \{1,2\}$$

Em que $U_j = 0$ e $V_k = 0$ significam a ausência de fótons, ou seja, o trajeto percorrido não foi o correspondente a $(W_1, W_2) = (j, k)$ (o outro trajeto que foi percorrido).

Portanto, a probabilidade condicional (dado que os fótons percorreram pelos trajetos relacionados com as variáveis aleatórias $(W_1, W_2)$ com valores iguais a $(w_1, w_2)$) será

$$\mathcal{P}_{U_1,U_2,V_1,V_2;W_1,W_2}(u_1, u_2, v_1, v_2; w_1, w_2) \equiv$$

$$\equiv \begin{cases} \mathcal{P}_{U_1,V_1}(u_1, v_1, \phi_{1,1}) \Leftarrow (w_1 = w_2 = 1) \wedge (u_2 = v_2 = 0) \\ \mathcal{P}_{U_1,V_2}(u_1, v_2, \phi_{1,2}) \Leftarrow (w_1 = w_2 - 1 = 1) \wedge (u_2 = v_1 = 0) \\ \mathcal{P}_{U_2,V_1}(u_2, v_1, \phi_{2,1}) \Leftarrow (w_1 - 1 = w_2 = 1) \wedge (u_1 = v_2 = 0) \\ \mathcal{P}_{U_2,V_2}(u_2, v_2, \phi_{2,2}) \Leftarrow (w_1 = w_2 = 2) \wedge (u_1 = v_1 = 0) \end{cases}$$

A esperança condicional (o sinal de ponto e vírgula separa as variáveis aleatórias $(U, V)$ das observadas $(W_1, W_2)$) do produto $U \cdot V$ será

$$\mathcal{E}_{U_1,U_2,V_1,V_2;W_1,W_2}(U_{w_1} \cdot V_{w_2}; w_1, w_2) \equiv$$

$$\equiv \sum_{U_{w_1}=-1}^{1} \sum_{V_{w_2}=-1}^{1} \left( U_{w_1} \cdot V_{w_2} \cdot \mathcal{P}_{U_1,U_2,V_1,V_2;W_1,W_2}(U_1, U_2, V_1, V_2; w_1, w_2) \right)$$

Portanto

$$\mathcal{E}_{U_1,U_2,V_1,V_2;W_1,W_2}(U_{w_1} \cdot V_{w_2}; w_1, w_2) \equiv \cos(2 \cdot \phi_{w_1,w_2})$$

Assim, com esse entendimento, a substituição feita na desigualdade seria proveniente não da esperança $\mathcal{E}_{U_1,U_2,V_1,V_2}$, mas sim da esperança condicional $\mathcal{E}_{U_1,U_2,V_1,V_2;W_1,W_2}$, portanto

$$cos(2 \cdot \phi_{1,1}) - cos(2 \cdot \phi_{1,2}) + cos(2 \cdot \phi_{2,1}) + cos(2 \cdot \phi_{2,2}) \equiv$$

$$\equiv \mathcal{E}_{U_1,U_2,V_1,V_2;W_1,W_2}(U_1 \cdot V_1; 1,1) - \mathcal{E}_{U_1,U_2,V_1,V_2;W_1,W_2}(U_1 \cdot V_2; 1,2) +$$

$$+ \mathcal{E}_{U_1,U_2,V_1,V_2;W_1,W_2}(U_2 \cdot V_1; 2,1) + \mathcal{E}_{U_1,U_2,V_1,V_2;W_1,W_2}(U_2 \cdot V_2; 2,2)$$

o que não faz muito sentido na Estatística misturar dessa forma esperanças condicionais em que os eventos dados são diferentes. Sabe-se que a probabilidade condicional de um dado evento ocorrido, possui as mesmas propriedades da probabilidade (demonstradas a partir dos axiomas de probabilidade), mas o evento ocorrido é mantido fixo.

O cálculo da esperança de $U_1 \cdot V_1$ com tal função de probabilidade seria dado por

$$\mathcal{E}_{U_1,U_2,V_1,V_2,W_1,W_2}(U_j \cdot V_k) \equiv \mathcal{E}_{W_1,W_2}\left(\mathcal{E}_{U_1,U_2,V_1,V_2;W_1,W_2}(U_j \cdot V_k; W_1, W_2)\right)$$

$$\mathcal{E}_{U_1,U_2,V_1,V_2;W_1,W_2}(U_j \cdot V_k; W_1, W_2) \equiv$$

$$\equiv \begin{cases} \mathcal{P}_{W_1}(W_1) \cdot \mathcal{P}_{W_2}(W_2) \cdot \cos(2 \cdot \phi_{W_1,W_2}) \Leftarrow ((j = W_1) \wedge (k = W_2)) \\ 0 \Leftarrow \neg((j = W_1) \wedge (k = W_2)) \end{cases}$$

$$\mathcal{E}_{U_1,U_2,V_1,V_2,W_1,W_2}(U_j \cdot V_k) \equiv \mathcal{P}_{W_1}(j) \cdot \mathcal{P}_{W_2}(k) \cdot \cos(2 \cdot \phi_{j,k})$$

Portanto a desigualdade seria

$$|(1-p_1) \cdot (1-p_2) \cdot \cos(2 \cdot \phi_{1,1}) - (1-p_1) \cdot p_2 \cdot \cos(2 \cdot \phi_{1,2}) + p_1 \cdot (1-p_2) \cdot$$
$$\cdot \cos(2 \cdot \phi_{2,1}) + p_1 \cdot p_2 \cdot \cos(2 \cdot \phi_{2,2})| \leqq 2$$

Esquecendo por um momento que um dos ângulos possuem uma dependência linear dos outros, atribuindo $\phi_{1,1} = \phi_{2,1} = \phi_{2,2} = 0$ e $\phi_{1,2} = \frac{\pi}{2}$, tem-se

$$|(1-p_1) \cdot (1-p_2) + (1-p_1) \cdot p_2 + p_1 \cdot (1-p_2) + p_1 \cdot p_2| \equiv$$
$$\equiv |(1-p_1) \cdot (1-p_2+p_2) + p_1 \cdot (1-p_2+p_2)| \equiv |(1-p_1) + p_1| \equiv 1 \leqq 2$$

Portanto, além de ser estatisticamente razoável, a desigualdade é matematicamente impossível de se violar, diferentemente se atribuirmos esses mesmos valores na desigualdade original

$$|cos(2 \cdot \phi_{1,1}) - cos(2 \cdot \phi_{1,2}) + cos(2 \cdot \phi_{2,1}) + cos(2 \cdot \phi_{2,2})| \equiv |1+1+1+1| \equiv 4$$

Mostrando que é matematicamente possível violar a desigualdade original.

## Conclusão

Devido a dificuldade encontrada na Física Quântica de se entender suas teorias, por não se encontrar análogos no Física Clássica, a falta de familiaridade e a multidisciplinariedade do mundo quântico tona difícil a aplicação coerente de todas as teorias envolvidas. Não se pode negar seu sucesso experimental, porém deve-se tomar o cuidado de observar se os dados obtidos no experimento obedecem às restrições teóricas da fórmula a ser usada. Neste caso, observou-se que usando a probabilidade condicional, a desigualdade se mantinha, quaisquer que fossem os parâmetros.

## Bibliografia

[1] ASPECT, A.; DALIBARD, J.; ROGER, G. Experimental test of Bell's inequalities using time-varying analyzers. **Physical Review Letters.** The American Physical Society. v. 49, n. 25, p. 1804-1807, 1982.

[2] MOOD, A. M.; GRAYBILL, F. A.; BOES, D. C. **Introduction to the theory of statistics**. 3rd. ed. New York: McGraw-Hill, 1973. 564p

[3] MANOUKIAN, E. B. **Quantum theory: a wide spectrum.** Dordrecht: Springer, 2006. 1011p

[4] PITOWSKY, I. **Quantum probability**: quantum logic. Berlin: Springer-Verlag, 1989. 209p

[5] KHRENNIKOV, A. I. U. **Contextual approach to quantum formalism.** New York: Springer, 2009. 353p